\documentstyle[12pt,agums]{article}

\lefthead{Summers and Ma}
\righthead{Relativistic electrons in the inner magnetosphere}

\received{???~??,~1999}
\revised {???~??,~1999}
\accepted{???~??,~1999}

\paperid{}
\cpright{AGU}{1998}
\ccc{0148-0227/96/96JA-00000\$09.00}
\authoraddr{Danny Summers and Chun-yu Ma,
Department of Mathematics and Statistics, 
Memorial University of Newfoundland, 
St. John's, Newfoundland, A1C 5S7, Canada
(e-mail: dsummers@math.mun.ca, cyma@math.mun.ca)}

\slugcomment{Submitted to {\it Journal of Geophysical Research
}, 1999.}

\begin{document}

\title{A MODEL FOR GENERATING RELATIVISTIC ELECTRONS IN THE
EARTH'S INNER MAGNETOSPHERE BASED ON GYRORESONANT WAVE-PARTICLE INTERACTIONS}

\author{Danny Summers and Chun-yu Ma \altaffilmark{1}}

\affil{Department of Mathematics and Statistics, Memorial University of 
Newfoundland,\\ St. John's, Newfoundland, A1C 5S7, Canada}

\altaffiltext{1}{
On leave from Purple Mountain Observatory, Chinese Academy of Sciences,
Nanjing, P.R. China.}

\begin{abstract} 
During the recovery phase of a magnetic storm, fluxes of relativistic ($>1$ 
MeV) electrons in the inner magnetosphere ($3\le L \le 6$) increase to 
beyond pre-storm levels, reaching a peak about 4 days after the initiation 
of the storm. In order to account for the generation of these ``killer 
electrons", a model is presented primarily based on stochastic 
acceleration of electrons by enhanced whistler-mode chorus. In terms of a 
quasi-linear formulation, a kinetic (Fokker-Planck) equation for the 
electron energy distribution is derived, comprising an energy diffusion 
coefficient based on gyroresonant electron-whistler-mode wave 
interaction and parallel wave propagation; a source term representing 
substorm-produced (lower 
energy) seed electrons; and a loss term representing electron precipitation 
due to pitch-angle scattering by whistler-mode waves and EMIC waves. 
Steady-state solutions for the electron energy distribution are 
constructed, and fitted to an empirically-derived relativistic Maxwellian 
distribution for the high energy ``hard" electron population at 
geosynchronous orbit. If the average whistler amplitude is sufficiently 
large, for instance 75 pT -- 400 pT, dependent on the values of the other 
model parameters, and assuming a background plasma density of $N_0=10$ 
cm$^{-3}$ outside the plasmasphere,
then a good fit to the empirical distribution is obtained, and 
corresponds to a timescale for the formation of the high-energy steady 
state distribution of 3 -- 5 days. 
For a lower representative value of the background plasma density, 
$N_0=1$ cm$^{-3}$, smaller whistler amplitudes, in the range 13 -- 72 pT, can
produce the high-energy distribution in the required time frame of several 
days. It is concluded from the model 
calculations that the process of stochastic acceleration by 
gyroresonant electron-whistler-mode wave interaction, in conjunction with 
pitch-angle scattering by EMIC waves, constitutes a viable mechanism for 
generating ``killer electrons" during geomagnetic storms.
The mechanism is expected to be particularly effective for the class of 
small and moderate storms possessing a long-lasting recovery phase during 
which many substorms occur.

\end{abstract}

\begin{article}

\section{INTRODUCTION}
It is well known that variations in the fluxes of relativistic electrons,
of kinetic energies $>$ 1 MeV, in the inner magnetosphere ($3\le
L\le 6$)
are related to disturbed magnetospheric conditions commonly called
``magnetic storms". Typically, for many storms the electron fluxes
diminish
rapidly during the main phase of the storm . The main phase depletion of
relativistic electrons occurs in association with large negative values of
the interplanetary magnetic field $B_z$ and large sudden increases in the
solar wind density and pressure \cite{pau, blake97}.
Subsequently, during the recovery phase of the storm, fluxes increase to
beyond pre-storm levels and peak about 4 days after the initiation of the
storm \cite{pau, baker86, nag, li1, ree}. 
These enhancements in fluxes of relativistic electrons,
which are colloquially referred to as ``killer electrons", have become the
subject of considerable attention by magnetospheric physicists. Not only
do the enhancements constitute an intrinsically interesting physics problem
in the near-Earth space, but they constitute a potentially serious hazard
to satellites, space stations, and, conceivably, humans in space. In fact,
satellite disfunctions (``anomalies") have been linked to the effects of
relativistic electron increases
\cite{baker94b},
and the state of the radiation belt environment  has become a major concern
in space weather forecasting [e.g.,
{\it Baker}, 1998; {\it Reeves}, 1998a].
The region near geosynchronous (or geostationary) orbit, $L\simeq 6.6$,
in the geographic equatorial plane, is of particular interest because it
is the operating zone of many orbiting satellites.
{\it Reeves} [1998b]
has recently examined the relationship between relativistic
electron enhancements at geosynchronous orbit and magnetic storms as
measured by the $D_{st}$ index. In particular, the 30 most intense
relativistic electron events during 1992-1995 were examined, and it was
found that every relativistic electron event was associated with a magnetic
storm as indicated by the $D_{st}$ index, though a small fraction (about
10\%) of magnetic storms did occur with no increase in relativistic
electron fluxes. Thus, one conclusion from {\it Reeves'} [1998b]
analysis is that intense solar wind conditions are necessary to
generate strong relativistic electron enhancements. Nevertheless,
despite the accumulated magnetic storm data from satellites over many
years, including coordinated observations from the International
Solar-Terrestrial Physics (ISTP) constellation of spacecraft and other
multi-satellite missions 
[e.g., {\it Baker et al.}, 1997; {\it Reeves et al.,} 1998],
there is, as yet, no accepted explanation for the generation of the
relativistic electrons. Specifically, it is not known exactly how, where,
or when the electrons are accelerated. Various energization mechanisms
have been proposed, and most of these are reviewed by {\it Li et al.}
[1997a]. It appears easier to explain the main phase depletion of energetic
electrons than their subsequent recovery and enhancement.
The drop in relativistic electron fluxes near geosynchronous orbit is
partly due to adiabatic responses (conserving all three adiabatic
invariants)
to magnetic field decreases, as reflected in the reduction in $D_{st}$
index [e.g., {\it Kim and Chan}, 1997]. Nevertheless, {\it Li et al.}
[1997a] show that other physical mechanisms, including precipitation,
must also contribute to the depletion. It has been suggested that
radial diffusion \cite{schulz},
invoked to explain  the existence of the outer electron radiation belt
itself, could also generate electrons of MeV energies in the inner
magnetosphere. This mechanism, which involves inwardly transporting
energetic electrons from a presumed source in the outer magnetosphere
(in the tail), can produce such energies during relatively quiet periods
[e.g., {\it Selesnick and Blake}, 1997a], although the process is too 
slow during active times
\cite{li1, blake98}.
Certain global recirculation processes, involving radial diffusion,
have also been proposed to generate relativistic electrons
[e.g., {\it Baker et al.}, 1986, 1989; {\it Fujimoto and Nishida}, 1990],
though these have proved inadequate, as the transport rates are too slow.
{\it Sheldon et al.} [1998]
have recently identified the outer polar cusp region as a potential
acceleration region of the magnetosphere and possible source of energetic
electrons for the outer radiation belt, though further calculations are
needed to evaluate the significance of the study.
In another mechanism still to be fully evaluated, {\it Rostoker et al.}
[1998] and {\it Liu et al.} [1999]
make the case that large-amplitude ULF pulsations have the
potential to supply the energy necessary to create the enhanced relativistic
electron fluxes. In a 3D global MHD simulation of the rapid rise of 
relativistic electron fluxes during the January 1997 magnetic cloud event,
{\it Hudson et al.} [1999a, b] also found that ULF oscillations may play 
a role in energizing relativistic electrons, via a mechanism involving 
drift-resonant acceleration and radial transport.

It is becoming increasingly apparent that electrons are accelerated to
relativistic ($>$ 1 MeV) energies {\it in situ} in the inner magnetosphere
[e.g., {\it Blake et al.}, 1998].
Significant evidence in support of this conclusion is the observation by
{\it Selesnick and Blake} [1997b]
that the phase space density of electrons of greater than MeV energies peaks
near L=4 to L=5 during storms. As a result of substorm activity,
electrons with energies up to $\sim 300$ keV are injected near
geosynchronous orbit
\cite{cay, baker89}.
These electrons appear to form the source population for the relativistic
electrons of greater than MeV energies that are subsequently observed.
{\it Summers et al.} [1998, 1999] have shown that whistler-mode waves could
provide an effective mechanism for accelerating electrons from energies near
100 keV to above 1 MeV in the region outside the plasmapause during the
storm recovery phase.
In a survey of potential wave modes for electron scattering and stochastic
acceleration to relativistic energies during magnetic storms,
{\it Horne and Thorne} [1998] concluded, in particular, that in low density
regions of the magnetosphere where the electron gyrofrequency exceeds the
electron plasma frequency, there are four potential wave modes that can
resonate with electrons in the energy range 100 keV to a few MeV:
the whistler, LO, RX, and Z modes. The concept of stochastic acceleration
of electrons by whistler-mode waves in the magnetosphere has also been
discussed by {\it Temerin et al.} [1994], {\it Li et al.} [1997a],
{\it Temerin} [1998], and {\it Roth et al.} [1999].
It is the purpose of present paper to quantify the model presented
qualitatively in Section 8 of {\it Summers et al.} [1998] for the
stochastic acceleration of relativistic electrons during geomagnetic
storms. Essential ingredients in the model are the spatial regions within
the inner magnetosphere $3\le L\le 9$ where enhanced whistler-mode chorus
\cite{tsu74, koons, par}
and enhanced electromagnetic ion cyclotron (EMIC or L-mode) waves occur
\cite{corn, per, jord,koz}; see Figure 1.
The aforementioned substorm-produced seed population of electrons in the
energy range 100 keV to 300 keV have approximately circular drift paths
within the region $3\le L \le 9$, and consequently will
traverse the regions of enhanced whistler-mode chorus and
enhanced EMIC waves. Specifically, in this paper we shall model the
acceleration of electrons during the storm recovery phase by means of
second-order Fermi (or stochastic gyroresonant) acceleration by weak
whistler-mode turbulence. In a standard quasi-linear formulation, we
construct a kinetic equation for the evolution of the electron energy
distribution function. The equation contains an energy diffusion
coefficient due to resonant whistler-mode wave/electron interaction,
and an electron loss term due to pitch-angle scattering by both the
whistler-mode and EMIC waves. We should point out here that the data
from the Solar, Anomalous, and Magnetospheric Particle Explorer (SAMPEX)
satellite \cite{nak95, li1, nak98}
show that bursty electron precipitation occurs as the electron flux
increases during the storm recovery phase. Such precipitation,
together with the existence of an abundant supply of storm-produced
lower-energy seed electrons
\cite{li1, baker98}
are supportive of the model constructed in this paper. The model
is presented in detail in Section 2, and numerical solutions are presented
in Section 3. The solutions are compared with data on the electron energy
distribution of high energy (300 -- 2000 keV) electrons at geosynchronous
orbit. We find that stochastic gyroresonant acceleration by whistler-mode
waves can indeed accelerate
substorm-produced seed electrons in the inner magnetosphere to generate
high-energy electron spectra of the type observed, following continuous
injection of seed electrons over a timescale of several days.
Specific predictions of the model depend, of course,
on the values taken for the model parameters.
In Section 4 we briefly assess our findings
and state our conclusions.

\section{MODEL}
The region to which the model constructed in this paper applies is that
part of the inner magnetosphere during storm-time that is illustrated
in the idealized 
\callout{Figure \ref{fig1}}, 
for $3\le L\le 9$. This region contains an
extensive subregion of whistler-mode chorus, a smaller but intense region
of EMIC waves, and also contains the important geosynchronous-orbit region
near L=6.6. Figure 1 is a simplified version of Figure 7 of {\it Summers
et al.} [1998], which itself was constructed on the basis of observations
and relevant theory
\cite{corn, per, koons, par, tsu74, koz, jord}.
We assume that geomagnetic storm activity produces a seed population of
electrons of energy $\approx 100$ keV as a source for the region specified
in Figure 1. We are not concerned in this paper with the precise means
(transport, original source location, etc.) by which the source is supplied.
According to standard particle drift theory [e.g., {\it Wolf}, 1995], the 
drift
motion of ``hot" (e.g., 100 keV) electrons close to the Earth is dominated
by gradient drift, with the result that electrons execute approximately
circular drift trajectories eastward about the Earth. Thus, the
storm-supplied source electrons constitute a quasi-trapped population
traversing the whistler-mode and EMIC wave subregions illustrated in
Figure ~\ref{fig1}. While executing the eastward drift, the electrons
gyrate about the field lines and `bounce' between mirror points, during
which time they also undergo both energy and pitch-angle diffusion, as a
result of their interaction with the whistler-mode chorus and EMIC waves.
We shall assume that the pitch-angle scattering rate is much greater than
the energy diffusion rate (see Appendix A), so that the electron
distribution will be nearly isotropic. Further, we shall take account of
pitch-angle diffusion of electrons into the loss-cone and their subsequent
precipitation into the atmosphere, by characterizing their loss from the
system by an escape time $T_{esc}$. The kinetic or Fokker-Planck equation
describing the evolution in time $t$ of the electron energy distribution
function $f(E,t)$ can be written as
\begin{equation}
  {\partial f\over {\partial t}}=
  {\partial^2\over{\partial E^2}}\left(D(E)f\right) -
  {\partial\over{\partial E}}\left[\left(A(E)-\mid\dot{E_L}\mid\right)f\right]-
  {f\over T_{esc}} + S(E,t),
 \label{eq1} 
\end{equation}
where $E=E_k/(m_ec^2)=\gamma - 1$ is the particle kinetic energy in units
of the rest mass energy, $\gamma=(1-v^2/c^2)^{-1/2}$ is the Lorentz factor,
$v$ is the particle speed, $m_e$ is the electron rest mass, and c is the
speed of light; $f(E,t) dE$ is the number of particles per unit volume in
the interval $dE$; $D(E)$ is the energy diffusion coefficient due to
resonant interaction of the electrons with whistler-mode turbulence;
$A(E)$ is the systematic acceleration rate due to the whistler-mode
turbulence; $\mid\dot{E_L}\mid$ is the energy loss rate due to processes
not
directly related to stochastic acceleration, namely, here assumed to be
Coulomb collisions and synchrotron radiation; $T_{esc}$ is the mean escape
time of particles out of the system due to pitch-angle scattering by both
whistler-mode and EMIC waves; and the source term $S(E,t)$ represents the
rate of particle injection into the inner magnetospheric region specified
in Figure ~\ref{fig1}, as a result of storm activity.
Equation (\ref{eq1}) is not the standard form of Fokker-Planck equation
employed in space physics, and so we give a brief account of its
derivation in the Appendix.
Detailed data on the whistler-mode chorus during storm-time are unfortunately
not available. In fact, insufficient information is known about the energy 
spectrum of the
turbulence in many space physics situations. While whistler-mode chorus
emissions are normally considered to be discrete during geomagnetically
quiet times \cite{anderson},
it is here assumed that during geomagnetic storms the  concomitant,
enhanced whistler-mode turbulence can be considered quasi-continuous.
Specifically, a simplifying assumption is made that the whistler-mode
turbulence is isotropic, homogeneous, stationary, and has a power-law
spectral energy density distribution in wavenumber $k$, with spectral
index $q$; specifically, the spectral energy density is assumed to take
the form,
\begin{equation}
W(k)={q-1\over{k_{min}}}\left({k_{min}\over k}\right)^qW_{tot},
~~~~~~
W_{tot}=\int_{k_{min}}^{\infty} {W(k) dk},
\end{equation}
for wavenumbers greater than $k_{min}$, to be specified below.
In accordance with the quasi-linear diffusion formulation adopted in this
paper, the whistler-mode turbulence is weak, i.e., comprises small-amplitude
magnetic and electric wave fields.
Momentum diffusion coefficients corresponding to whistler-mode waves
have been obtained by various authors. We calculate the Fokker-Planck
coefficients $D(E)$ and $A(E)$ from (A8) using the whistler-mode
diffusion coefficients $D_p$ derived by
{\it Hamilton and Petrosian} [1992] (for $2< q\le 4$), and
{\it Schlickeiser} [1997] (for $1< q < 2$) for parallel wave propagation.
The results are
\begin{equation}
D(E)=D_0\left[E(E+2)\right]^{(q-1)/2}(E+1)^{-1},
\end{equation}
\begin{equation}
A(E)=D_0 q \left[E(E+2)\right]^{(q-3)/2},
\end{equation}
where
\begin{equation}
D_0={\pi(q-1)^2\over{q^2(q^2-4)}}
    \left({ck_{min}\over{\Omega_e}}\right)^{q-1}
    \alpha^2 R \Omega_e,
\end{equation}
for $2< q\le 4$;
\begin{equation}
D(E)={\mathcal D}_0\left[E(E+2)\right]^{1/2}(E+1)^{-1},
\end{equation}
\begin{equation}
A(E)=2 {\mathcal D}_0 \left[E(E+2)\right]^{-1/2},
\end{equation}
where
\begin{equation}
{\mathcal D}_0={\pi(q-1)\over 8}
    \left({ck_{min}\over{\Omega_e}}\right)^{q-1}
    \left({m_e\over m_p}\right)^{(2-q)/2}
    \alpha^{(2+q)/2} J_W R \Omega_e,
\end{equation}
for $1 < q < 2$.\\
In (3) -- (8), the two dimensionless parameters $R$ and $\alpha$ are 
introduced;
$R$ is the ratio of turbulent energy $W_{tot}$ to magnetic field energy,
\begin{equation}
R=8\pi W_{tot}/B_0^2=\left(\Delta B/B_0\right)^2,
\end{equation}
and
\begin{equation}
\alpha=\Omega_e^2/\omega_{pe}^2=(m_p/m_e)\beta_A^2,
\end{equation}
where $\Omega_e=eB_0/(m_ec)$ is the electron gyrofrequency, with $B_0$
the ambient magnetic field strength and $e$ the electron charge;
$\Delta B$ is the average whistler-mode wave amplitude;
$\omega_{pe}=(4\pi N_0e^2/m_e)^{1/2}$ is the electron plasma frequency, with
$N_0$ the particle number density; $J_W$ is a weakly varying function of $E$;
and $\beta_A=v_A/c$ where $v_A=B_0/(4\pi N_0m_p)^{1/2}$ is the
Alfv\'en speed, with $m_p$ the proton rest mass. The parameter $\alpha$
defined in (10) is identical to the parameter $\alpha$ used by
{\it Summers et al.} [1998].
We note, in particular, as should indeed be the case, that since the kinetic
energy variable $E$ is dimensionless, the dimension of the diffusion
coefficient $D$ equals the dimension of the parameter $D_0$ 
(or ${\mathcal D}_0$)
equals [time]$^{-1}$. For definiteness, in (5) and (8) we set
\begin{equation}
k_{min}=\Omega_p/(c\beta_A),
\end{equation}
[e.g., {\it Hamilton and Petrosian}, 1992], where $\Omega_p=eB_0/(m_pc)$
is the proton gyrofrequency.

The energy loss term $\mid\dot{E_L}\mid$, in which we include losses due
to Coulomb collisions and synchrotron radiation, can be expressed in the
form
\begin{equation}
\mid\dot{E_L}\mid=6\times 10^{-13} N_0 (E+1)[E(E+2)]^{-1/2}+
         1.32\times 10^{-9} B_0^2 E(E+2).
\end{equation}
The first term on the right-hand side of the equation (12) is the energy
loss rate due to Coulomb collisions, given by
{\it Melrose} [1980],
and the second term the energy loss rate due to synchrotron radiation, 
given by {\it Blumenthal and Gould} [1970].
In (12), the particle number density $N_0$ is in cm$^{-3}$, the magnetic
field strength $B_0$ is in gauss, and $\mid\dot{E_L}\mid$ is in
sec$^{-1}$.

We note that there are potentially four influential parameters in the
model: the parameter $\alpha$ defined by (10), the spectral index $q$, the
turbulent wave power parameter $R$, and the mean particle escape-time
$T_{esc}$. The value of the parameter $\alpha$ depends on the values taken
for the particle density $N_0$ and the ambient magnetic field strength
$B_0$. We shall discuss $T_{esc}$, which we regard as an adjustable
parameter, and the particle source function $S$ below. The diffusion
parameters $D_0$ and ${\mathcal D}_0$ occurring in expressions (3) and (6)
for the diffusion coefficient $D$ are measures of the rate of energy
diffusion, and $D_0^{-1}$, ${\mathcal D}_0^{-1}$ are measures of the time
scale for particle acceleration. Substituting the result (11) for
$k_{min}$ into equations (5) and (8), we find that $D_0$ and ${\mathcal
D}_0$ are given by

\begin{equation}
D_0={\pi(q-1)^2\over{q^2(q^2-4)}}
    \left({m_e\over{m_p}}\right)^{q-3}
    \Omega_e R \beta_A^{5-q}
   ={\pi(q-1)^2\over{q^2(q^2-4)}}
    \left({m_e\over{m_p}}\right)^{(q-1)/2}
    \Omega_e R \alpha^{(5-q)/2},
\end{equation}
for $2< q\le 4$;
\begin{equation}
{\mathcal D}_0={\pi(q-1)\over 8}
    \left({m_p\over{m_e}}\right)
    J_W \Omega_e R \beta_A^3
    ={\pi(q-1)\over 8}
    \left({m_e\over{m_p}}\right)^{1/2}
    J_W \Omega_e R \alpha^{3/2},
\end{equation}
for $1 < q < 2$.\\
Corresponding to the Kolmogorov turbulent spectrum ($q=5/3$), the function
$J_W$ is of order unity. 
As an idealized assumption, we regard the Kolmogorov spectrum as the 
representative spectrum over the range $1 < q < 2$,
and we henceforth set $q=5/3$ and $J_W=1$ in (14).

Since, from (10), the parameter $\alpha$ is inversely proportional to the
particle number density $N_0$, it follows from (13) and (14) that $D_0$
and
${\mathcal D}_0$ increase as $N_0$ decreases. This agrees with the
conclusions of
{\it Summers et al.} [1998]
who found by constructing resonant diffusion curves in velocity space that
energy diffusion becomes more pronounced with increasing $\alpha$
(or decreasing $N_0$). As expected, the values of $D_0$ and ${\mathcal
D}_0$ also increase as the turbulent spectral energy density ratio $R$
increases.
Specifically, we find from (13) and (14) that $D_0$ and ${\mathcal D}_0$
depend on the plasma parameters $N_0$ and $B_0$, and the wave amplitude
$\Delta B$ as follows:
\begin{equation}
D_0 \propto B_0^{(4-q)}(\Delta B)^2/N_0^{(5-q)/2}
\end{equation}
for $2<q\le4$;
\begin{equation}
{\mathcal D}_0 \propto B_0^2(\Delta B)^2/N_0^{3/2}
\end{equation}
for $1<q<2$.

In this paper, we set $N_0=10$ cm$^{-3}$ as the particle number density
representative of the inner magnetosphere ($3\le L \le 9$) outside the
plasmasphere.  It could be argued that such a value may be too high for the
background plasma outside the plasmasphere. 
However, since from (15) and (16) it is clear that the acceleration 
process becomes more efficient as $N_0$ decreases, we find it useful to 
adopt $N_0=10$ cm$^{-3}$ as a generic conservative value. 
We comment further on this assumption below. 
We use the equatorial (dipole)
magnetic field value $B_0=3.12\times 10^{-5}/L^3$ T. Corresponding values
of $B_0$ and the above-defined parameters $\alpha$ and $\beta_A$ at the
locations $L = 3,$ 4, $\cdots$, 9 are given in Table 1. 
In \callout{Figure \ref{fig2}}, we plot the energy diffusion parameter
$D_0$ (sec$^{-1}$) as a function of the spectral index $q$ in the range
$2<q<4$ at each of the locations $L=3$, 4, 5. At each $L$-value, we 
calculate
$D_0$ for the specified wave amplitudes $\Delta B=75$ pT, 100 pT, 300 pT,
and 1 nT ( which correspond to the indicated values of $R$ in the diagrams).
Lines indicating the time scales $D_0^{-1}$ for particle acceleration
corresponding to 1 hour, 1/2 day, and 1 day are shown in each diagram.
As can be observed from the curves in Figure 2, the value of $D_0$ is
particularly sensitive to the value of $q$ as $q$ approaches 2. In fact,
formally from (13) we have the result $D_0\rightarrow \infty$, as 
$q\rightarrow 2$, 
which is obviously undesirable physically, but which is a consequence of
the quasi-linear diffusion formalism we have adopted in this paper.
It is evident from Figure 2 that, at any given $L$-value, as the value of $q$
decreases, the value of $D_0$ increases, and hence the time scale for
particle acceleration decreases.  In addition, it can be observed that for
a given value of $q$, as $L$ decreases, the value of $D_0$ likewise 
increases;
this property also follows from relation (15). Thus, for $q$ in the range
$2<q<4$, shorter acceleration times are favoured by smaller value of $q$,
small values of $L$, and (of course) larger values of the wave amplitude
$\Delta B$. Corresponding to the Kolmogorov spectrum ($q=5/3$), we show
in \callout{Figure \ref{fig3}} the
diffusion parameter ${\mathcal D}_0$ (sec$^{-1}$) as a function
of $\Delta B$ (pT) at the locations $L=3$, 4, 5. Again, it is clear from the 
figure
that shorter acceleration times are favoured  by smaller values of $L$ and
larger values of $\Delta B$; this property similarly follows from (16).

In \callout{Figure \ref{fig4}}, 
we plot the diffusion coefficient $D$ (sec$^{-1}$) as a function
of the particle kinetic energy $E$ (MeV), as given by (3), (6), (13) and (14),
for the fixed wave amplitude $\Delta B=1$ nT, and $N_0=10$ cm$^{-3}$.
The curves are constructed for values of the spectral index $q=5/3$, 2.5, 3,
3.5, and 4, at each of the locations, $L=3$, 4, 5. The diffusion coefficient
$D$ is clearly an increasing function of energy $E$. In general, for a given 
value of $E$, though not at all values, $D$ can also be seen to increase
as $q$ decreases.

According to the standard quasi-linear theory of resonant interaction of
electrons with whistler-mode turbulence [e.g., {\it Melrose}, 1986], in 
order for electrons to resonate with whistlers, the condition,
\begin{equation}
\gamma\beta \ge (m_p/m_e)^{1/2}\beta_A
\end{equation}
must be satisfied; $\beta=v/c$ where $v$ is the particle speed, $c$ is the
speed of light, $\gamma$ is the Lorentz factor, and $\beta_A=v_A/c$ is the
Alfv\'en speed parameter defined above. Making use of relativistic relations
given in (A6), we find that (17) can be expressed in the form
\begin{equation}
E(E+2) \ge (m_p/m_e)\beta_A^2
\end{equation}
which, in turn, by using the parameter $\alpha$ defined in (10), can be
reduced to
\begin{equation}
E\ge E_c,
\end{equation}
where $E_c$ is the critical energy given by
\begin{equation}
E_c=(1+\alpha)^{1/2}-1.
\end{equation}
The value of the parameter $\alpha$ depends on the values of the particle
number density $N_0$ and magnetic field $B_0$. Values of the critical
energy $E_c$ are given in \callout{Table 1} at the locations $L=3$, 4,
$\cdots$, 9, for $N_0=10$ cm$^{-3}$; the values for $L=3$, 4, 5 correspond
respectively to the energy cut-off values in the upper, middle, and lower
diagrams in Figure 4. 

\section{NUMERICAL RESULTS}
Prior to consideration of the solution of the kinetic equation (1) for the
electron energy distribution function $f(E,t)$, we must specify the source
function $S(E,t)$ which represents storm-produced seed electrons.
We shall assume that the source function can be represented by a standard
relativistic Maxwellian distribution, namely,
\begin{equation}
S=S_0[\mu/K_2(\mu)](E+1)[E(E+2)]^{1/2}e^{-\mu (E+1)}
\end{equation}
where
\begin{equation}
\mu=m_ec^2/(k_BT_s)
\end{equation}
and $T_s$ represents the temperature of the distribution; $K_2(\mu)$ is
a modified Bessel function of the second kind of argument $\mu$, and
$k_B$ is Boltzmann's constant. It can be shown that 
$\int{S}dE = S_0,$
so that the parameter $S_0$ represents the total number of source electrons
per unit volume per unit time.

It is clearly important both to calibrate and test our model, as far as is
possible at the present time, by making use of available observational data.
The study by {\it Cayton et al.} [1989] appears best-suited to these 
purposes.
{\it Cayton et al.} [1989] derived energy distribution functions from
energetic (30 -- 2000 keV) electron fluxes observed simultaneously by three
satellites in geosynchronous orbit throughout the year 1986. It was
found that the energetic electron population can be resolved into two
distinct relativistic Maxwellian components, each parameterized by a
temperature and a density: a lower energy (30 -- 300 keV) ``soft" electron
distribution  with temperature $T_s\approx 25$ keV and number density
$N_s\approx 5\times 10^{-3}$ cm$^{-3}$; and a higher energy (300 -- 2000 keV)
``hard" electron distribution with temperature $T_h\approx 200$ keV and 
number
density $N_h\approx 10^{-4}$cm$^{-3}$. The ``soft" component is charaterized
by intense substorm-related injections and by strong temporal variations.
Accordingly, and in agreement with a suggested interpretation by
{\it Cayton et al.} [1989], we shall regard this ``soft" component as
comprising the electron seed population. Thus, we shall identify the temperature
$T_s$ associated with the Maxwellian source distribution (21) -- (22)
as the aforementioned temperature of the ``soft" electron component.
{\it Cayton et al.} [1989] found that the value of $T_h$ shows 
little change
on the substorm (hourly) time scale, while $N_h$ decreases during substorms.
We shall regard the ``hard" electron distribution as precisely the highly
energetic (``killer") electron distribution that we are trying to model
as a (steady-state) solution of the kinetic equation (1) with the steady
Maxwellian source (21) -- (22).

We solve equation (1) for the energetic electron distribution $f(E,t)$
by the Crank-Nicholson implicit differencing scheme. The method is well 
suited to time-dependent Fokker-Planck equations, and we refer the reader 
to {\it Hamilton et al.} [1990] and {\it Park and Petrosian} [1996] for 
full details. Since we are concerned with the generation of a 
highly-energetic electron distribution, we assume that there are no 
such energetic particles initially, i.e., 
\begin{equation}
f(E,0)=0, ~~~~~~~~~~~~~~~~~~E>E_s,
\end{equation}
where $E_s=1/\mu$ is the thermal energy associated with the source 
distribution (21) -- (22). We further assume that, subject to continuous  
injection of the seed electrons (given by (21) -- (22)), the evolving 
distribution maintains a maximum at $E=E_s$ for all time, i.e., we take 
the inner boundary condition as
\begin{equation}
\frac{\partial f(E,t)}{\partial E} = 0, ~~~~~~~~~~~~~~ E=E_s.
\end{equation}
Finally, for the outer boundary condition we require that the 
distribution function tend to zero for large values of $E$ for all time, 
so we set
\begin{equation}
f(E,t)=0, ~~~~~~~~~~~~~~~~E>E_0
\end{equation}
where $E_0$ is a specified upper value of $E$ (in practice, we fix 
$E_0=2\times 10^4$ MeV).\\
Having constructed the evolving electron distribution subject to the 
above conditions, we thereby obtain the resulting steady-state 
distribution $f(E)$ which we fit to a relativistic Maxwellian 
distribution, i.e., we carry out the linear fit,
\begin{equation}
\log_{10}\left[f(E)/\{(E+1)[E(E+2)]^{1/2}\}\right ] \equiv a+bE,
\end{equation}
with
\begin{equation}
a=\log_{10}\left[N_h\lambda e^{-\lambda}/K_2(\lambda)\right], 
~~~~~~~~~~~~ b=-\lambda\log_{10}e,
\end{equation}
and
\begin{equation}
\lambda=m_ec^2/(k_BT_h),
\end{equation}
where $T_h$ and $N_h$ represent respectively the temperature and number 
density of the steady-state distribution (to be compared with the above 
values associated with the ``hard" electron distribution); and 
$K_2(\lambda)$ is a modified Bessel function of the second kind of 
argument $\lambda$. For a given steady-state solution, the parameters 
$a$ and $b$ are determined by a linear regression comprising a 
minimization of a chi-square goodness-of-fit merit function [{\it Press et 
al.}, 1992]. Having thus obtained values for $a$ and $b$, we then 
calculate $T_h$ and $N_h$ from (27) and (28).

Representative numerical solutions of the model presented in this paper 
are shown in \callout{Figures \ref{fig5}, \ref{fig6}}, and
\callout{\ref{fig7}}, 
with corresponding results given 
respectively in 
\callout{Tables 2, 3}, and \callout{4}. In 
all cases we set the background 
number density $N_0=10$ cm$^{-3}$, and the source electron temperature 
$T_s=25$ keV (giving $\mu=20.44$ from (22)), the latter value being 
equal to the estimate by
{\it Cayton et al.} [1989] for the temperature 
of the ``soft" electron distribution. The scheme and rationale for 
setting the remaining parameters is as follows. Firstly, we set $L$, which 
fixes the value of the background magnetic field $B_0$. Secondly, 
we set the average wave-amplitude $\Delta B$ and the spectral index $q$; 
these values are chosen to yield a value of the diffusion parameter $D_0$ 
(or ${\mathcal D}_0$) 
that is expected to 
produce a steady-state (equilibrium) distribution function after several 
days of source injection. Whistler-mode ``chorus" wave amplitudes 
have been reported in the range  1 -- 100 pT
[{\it Burtis and Helliwell,} 1975], with 
{\it Parrot and Gaye} [1994] finding that 
during more intense periods of magnetic activity
wave amplitudes can approach $\Delta B=1$ nT. 
Amplitudes of whistlers associated mainly with hiss, with values of 100 
pT or more, have also been reported by {\it Smith et al.} [1974]
during  a typical storm recovery phase.
In Table 1, corresponding to a background number density of $N_0=10$ 
cm$^{-3}$, we present the wave amplitudes $\Delta B$ (pT) expected to 
yield a high energy ``hard" electron distribution after a few days of 
seed electron injection as a result of substorm activity. These required 
amplitudes depend on the value of $L$ and $q$ (as well as $N_0$). For 
$q=5/3$, and for the inner region $3\le L \le 5$, the values of $\Delta 
B$ are in the range 75 -- 400 pT, which are realistic though in the 
higher range of observations. As pointed  out in Section 2, the process 
of gyroresonant stochastic acceleration becomes more efficient with 
decreasing background plasma density. While $N_0=10$ cm$^{-3}$ can be 
regarded as a representative value for the background plasma density 
outside the plasmasphere in certain conditions, it is also true that at 
other times $N_0=1$ cm$^{-3}$ is a more representative value. Using (15), 
we calculate that if we set $N_0=1$ cm$^{-3}$, then, for $3\le L \le 5$, 
the required $\Delta B$-values  are in the range 13 -- 72 pT, if $q=5/3$, 
and in the range 35 -- 316 pT, if $2.5\le q \le 3$.
Having set values for $N_0, T_s, L, q,$ and $\Delta B$, we next 
specify the mean particle escape time $T_{esc}$. In fact, since a value 
for $T_{esc}$ is not precisely known, we treat $T_{esc}$ as an adjustable 
parameter and run the cases $1/(D_0T_{esc})$ (or $1/({\mathcal D}_0T_{esc})) 
=0, 1, 2, 5, 10$. In Figure 5 (left) we show steady-state solutions for 
the electron energy distribution function $f(E)$ for the case $L=3$, 
$q=5/3$, 
$\Delta B=75$ pT. The source strength $S_0$ has been chosen so as to 
produce a model solution that best agrees with the  
``hard" electron distribution of {\it Cayton et al.} [1989]. In order to 
achieve this, for each 
steady-state solution the linear fit (26) -- (28) to a relativistic 
Maxwellian distribution is carried out. The corresponding results are 
shown in Figure \ref{fig5} (right) and 
Table 2. It is found that the temperature 
$T_h$ associated with a particular steady-state solution, as given by the 
parameter $b$ (or the slope of the constructed line), is largely 
determined by the value of $T_{esc}$, while the number density $N_h$, 
which is then given by the parameter $a$ (or the vertical intercept of 
the line) is largely determined by the value of $S_0$. The results 
in Table 2, 
for which $S_0=1.5\times 10^{-6}$ cm$^{-3}$sec$^{-1}$, indicate a best 
agreement with the empirically-derived
values of $N_h\approx 10^{-4}$ 
cm$^{-3}$ and $T_h=200$ keV, when $T_{esc}\approx 1/2$ day, the 
corresponding time for the formation of the steady-state solution being 
$T_{EQ}\approx 4$ days. A precise, physically representative value for the 
mean particle escape time $T_{esc}$ is difficult to determine {\it a 
priori} since $T_{esc}$ relates to scattering losses of electrons due to 
both whistler-mode and EMIC waves. However, on the basis of estimates of 
timescales for strong diffusion scattering loss, it appears that 
$T_{esc}$ is of the order of hours and so 
$T_{esc}\approx 1/2$ day is not an unreasonable value. We relate $T_{EQ}$ 
to the time taken after the initiation of a storm for fluxes of 
relativistic electrons to peak (see Section 1), which is observed to be 
several days. Thus, we favour solutions of the present model for which 
$T_{EQ}=$ 1 -- 5 days, with $T_{EQ}\approx 4$ days possibly the optimal 
value. 

In Figure \ref{fig6} and 
Table 3 
we show the corresponding results for the case $N_0=10$ cm$^{-3}$,
$L=6.6$, $q=2.5$, $\Delta B= 800$ pT,  while in Figure \ref{fig7} and 
Table 4 
we show the results for the case $N_0=10$ cm$^{-3}$, $L=5$, $q=3$, $\Delta 
B= 1$ nT. As can be seen from the 
tables, for both these cases best agreement between the solutions and 
the ``hard" electron distribution of {\it Cayton et al.} [1989] occurs when 
$T_{esc}\approx 1/2$ day, and corresponds to a formation time $T_{EQ}=$
3 -- 5 days.
If the background number density is taken to be $N_0=1$ cm$^{-3}$, the 
cases shown in Figures 6 and 7 correspond respectively to values for the 
wave amplitude $\Delta B$ of 190 pT and 316 pT. Figures 6 and 7 correspond 
to cases of intense substorm activity during the storm recovery phase.

Taking into account the value of the wave amplitudes given in Table 2 
corresponding to $N_0=10$ cm$^{-3}$, and their converted values for the 
case $N_0=1$ cm$^{-3}$, we re-iterate that the model solutions imply in 
particular that for a Kolmogorov turbulent wave spectrum, sustained 
whistler amplitudes in the physically realistic range 13 -- 72 pT can 
generate a typical high energy ``hard" electron distribution in the inner 
magnetosphere $3\le L\le 5$ within one or two days. It should also be noted 
that the model 
calculations show that the acceleration mechanism considered in this 
paper is not effective in the region $7\le L\le 9$ since the necessary 
values of the whistler amplitude would be too high (typically in excess 
of 1 nT). Thus, the model formulated herein has been shown to be a viable 
mechanism for accelerating electrons exactly in the inner region of the 
magnetosphere where the peak in electron phase space density of the 
highly energetic electrons is observed to occur [e.g., {\it Selesnick and 
Blake}, 1997b]. 

A requirement of the model presented here is enhanced 
whistler-mode chorus lasting for a period of at least one or two days. 
Geomagnetic conditions during which such a requirement is particularly 
well satisfied occur during the descending phase of the solar cycle when 
the Earth's magnetosphere can be impacted by a high-speed solar wind stream  
following a magnetic field build-up known as a Corotating 
Interaction Region (CIR). CIRs cause small and moderate magnetic storms, but 
not major storms. Since the Earth can be embedded in the associated 
high-speed stream for days to weeks there are substorms for days to weeks 
[{\it Tsurutani et al.}, 1995; {\it Kamide et al.}, 1998]. Thus, during 
this long-lasting recovery phase of the magnetic storm, there 
will be continuously enhanced wave activity, in the form of both 
whistler-mode chorus and EMIC waves, to drive the acceleration mechanism 
presented herein to generate the high energy ($>$1 MeV) electrons.
 
\section{CONCLUSIONS}

In this paper, by means of quasi-linear theory and a test-particle
approach, we have formulated the model kinetic equation (1) in which the
acceleration mechanism is due to gyroresonant interaction between
electrons and whistler-mode turbulence, corresponding to parallel wave
propagation.  The essential purpose of the study has been to apply
equation (1) to the Earth's inner magnetosphere in order to test the
hypothesis that storm-enhanced whistler-mode chorus can accelerate
lower-energy substorm-produced seed electrons to relativistic ($>1$ MeV)
energies over a period of a few days. Our conclusions are as follows: 
\begin{enumerate}
\item
Based on the model calculations in this paper, it is entirely possible for
enhanced whistler-mode chorus to generate the observed increases in
relativistic ($>$ 1 MeV) ``killer" electrons during the storm recovery
after a period of several days, so long as the waves are sufficiently
strong.  If $N_0=10$ cm$^{-3}$ is taken to be the background plasma
number density outside the plasmasphere, the typical average
wave-amplitudes required for a Kolmogorov spectrum are in the range
$\Delta B=75$ -- 400 pT, dependent on the location $L$. If $N_0=1$ 
cm$^{-3}$, the required wave-amplitudes are in range 13 - 72 pT.
\item
Energetic electron spectra of the model solutions do not follow a
simple power law in energy. For certain sets of parameters, we find that
solutions can be well fitted to the relativistic Maxwellian distribution
empirically constructed by {\it Cayton et al.} [1989] to represent the
higher energy (300 keV -- 2 MeV) ``hard" electron population at
geosynchronous orbit. We note the recent analysis by {\it Freeman et al.}
[1998] of the November 3 -- 4, 1993 storm, in which electrons from about 100
keV to 1.5 MeV were characterized by a power law spectrum. Evidently,
optimal fitting of empirical electron spectra to power-law, Maxwellian,
or other types of distribution can depend on the energy range prescribed
and the event under consideration. In connection with electron power-law
energy spectra, {\it Ma and Summers} [1998] have shown that such spectra
can be produced by whistler-mode turbulence, although it is questionable 
whether the necessary
conditions established in their theoretical study can be
satisfied in the Earth's magnetosphere.
\item
It is unlikely that any single physical mechanism of electron acceleration
can fully account for relativistic electron enhancements occurring during
the recovery phase of magnetic storms, not least because various types of
energetic electron event have been observed [e.g., {\it Baker et al.}, 
1997, 1998; {\it Reeves}, 1998b; {\it Reeves et al.}, 1998].
Rapid energetic electron flux enhancements taking place over minutes have 
been associated with inductive electric fields [e.g., {\it Li et al.}, 
1993], while enhancements occurring over tens of minutes or a few hours 
have been linked to ULF pulsations [e.g., {\it Rostoker et al.}, 1998; 
{\it Liu et al.}, 1999]. The gradual acceleration process (occurring over 
a few days) formulated in this paper is not intended to apply to such 
energetic electron events which typically result from major storms.
However, small and moderate magnetic storms associated 
with Corotating Interaction Regions (CIRs) characteristically have 
long recovery phases and attendant substorms for days to weeks [{\it 
Tsurutani et al.}, 1995; {\it Kamide et al.}, 1998]. Since these 
substorms produce enhanced whistler-mode chorus (and EMIC waves) over (at 
least) several days, the necessary conditions for the effectiveness of 
the mechanism presented in this paper are satisfied. Hence, for 
these types of storm, and possibly 
others, when average wave-amplitudes are sufficiently large, the 
present study shows that, in conjunction with pitch-angle scattering by 
EMIC waves, the mechanism of stochastic acceleration by whistler-mode
turbulence is a serious candidate for explaining the generation of
``killer electrons".
\end{enumerate}

\appendix

\section{Appendix A: Derivation of the Fokker-Planck equation (1)}

We consider energetic charged particles in a uniform magnetic field,
with superimposed small-amplitude plasma waves of a given mode. The
equation for the  evolution of the particle distribution function
$\phi(p,t,\mu)$ due to gyroresonant interactions with the waves is
 \begin{eqnarray}
 {{\partial \phi}\over{\partial t}} & = & 
{1\over p^2}{{\partial }\over{\partial p}}\left(p^2D_{pp}{{\partial 
\phi}\over{\partial p}}\right) +
{1\over p^2}{{\partial }\over{\partial p}}\left(p^2D_{p\mu}{{\partial 
\phi}\over{\partial \mu}}\right) +
 {{\partial }\over{\partial \mu}}\left(D_{\mu p}{{\partial 
\phi}\over{\partial p}}\right)\cr
 & & +
{{\partial }\over{\partial \mu}}\left(D_{\mu\mu}{{\partial 
\phi}\over{\partial \mu}}\right) 
 +{1\over{p^2}}{{\partial }\over{\partial 
p}}\left(p^2\dot{\mathcal{E}}_L\phi\right)
 +Q(p,t).
\label{A1}
 \end{eqnarray} 
Equation (\ref{A1}), called a kinetic or diffusion or
Fokker-Planck equation, is derived by expanding a collisionless Boltzmann
equation for the particle distribution function to second order in
perturbed quantities, and ensemble-averaging over the statistical
properties of the plasma waves in accordance with quasi-linear theory.
Among the early authors to carry out this procedure were {\it Kennel and
Engelmann} [1966], {\it Hall and Sturrock} [1967], and {\it Lerche}
[1968]; see also {\it Melrose} [1980], {\it Schlickeiser} [1989], and {\it
Steinacker and Miller} [1992], and references therein. In (\ref{A1}), $p$
is the relativistic unit momentum given by $p=\gamma v/c$, where $v$ is
the particle speed, and $\gamma=(1-v^2/c^2)^{-1/2}$ is the Lorentz factor,
with $c$ the speed of light; $t$ is time; $\mu$ is the cosine of the pitch
angle; $\dot{\mathcal{E}}_L$ is an energy loss term due to processes not 
directly
associated with gyro-resonant wave-particle interactions; and $Q(p,t)$ is
a source term. The Fokker-Planck or diffusion coefficients $D_{pp},
D_{p\mu}, D_{\mu p}$, and $D_{\mu\mu}$ depend on the properties of the
wave turbulence, viz., the wave mode and polarization, the angle of wave
propagation to the ambient magnetic field, and the power spectrum,
including the ratio of the turbulent wave energy to the background
magnetic energy. These coefficients have been given both in general form,
and specific form, for various particular wave modes, by a number of
authors, e.g., {\it Melrose} [1980], {\it Schlickeiser} [1989], {\it
Steinacker and Miller} [1992], and {\it Hamilton and Petrosian} [1992]. It
is not necessary here to derive equation (\ref{A1}), which requires
considerable algebra, or to provide general expressions for the
coefficients $D_{pp}, D_{p\mu}=D_{\mu p}$, and $D_{\mu\mu}$. We shall
assume that the rate of pitch-angle scattering is much larger than the
rate of energy diffusion (and the rate of the particle escape from the
system). Such an assumption is reasonable based on an analysis of time
scales associated with resonant interaction of electrons with
whistler-mode waves, e.g., see {\it Melrose} [1980] and the discussion by
{\it Hamilton and Petrosian} [1992]. Equivalently, defining the time
scales $T_{\mu\mu}=D_{\mu\mu}^{-1}$, $T_{\mu p}=pD_{\mu p}^{-1}$,
$T_{pp}=p^2D_{pp}^{-1}$, and the escape time $T_{esc}$, we assume that
$T_{\mu\mu}\ll T_{pp}$, $T_{\mu\mu}\ll T_{\mu p}$, and $T_{\mu\mu}\ll
T_{esc}$. Then the particle distribution function can be assumed to be
isotropic, and the pitch angle can be eliminated from the problem by
integrating (\ref{A1}) with respect to $\mu$ [e.g., see {\it
Schlickeiser}, 1989; {\it Steinacker and Miller}, 1992]. Writing
\begin{equation} F(p,t)=\int_{-1}^{1}{\phi(p,t,\mu)d\mu}, \label{A2}
\end{equation} and representing the scattering loss of particles by pitch
angle diffusion by means of a loss term $-F(p,t)/T_{esc}$, the equation
(\ref{A1}) thus becomes
 \begin{eqnarray}
 {{\partial F(p,t)}\over{\partial t}} & =& 
{1\over p^2}{{\partial }\over{\partial p}}\left(p^2D_p(p){{\partial 
 F(p,t)}\over{\partial p}}\right)+
 {1\over{p^2}}{{\partial }\over{\partial p}}\left(p^2\dot{\mathcal{E}}_L(p)
F(p,t)\right)\cr
& &
-{F(p,t)\over T_{esc}}+{1\over 2}Q(p,t),
\label{A3}
 \end{eqnarray}
where the momentum diffusion coefficient $D_p(p)$ has been formed by
averaging with respect to $\mu$.

We now change the momentum variable $p$ to the kinetic energy variable
$E=\gamma-1$ in equation (\ref{A3}). We write
\begin{equation}
f(E,t)dE=4\pi p^2F(p,t)dp,
\end{equation}
\begin{equation}
{\partial\over{\partial p}}={dE\over{dp}}{\partial\over{\partial E}},
\end{equation}
and make note of the following simple relativistic relations:
\begin{eqnarray}
&p=\gamma\beta, \beta=v/c, p^2=E(E+2), \gamma=(1+p^2)^{1/2},\cr
&pdp=(E+1)dE, \beta dp=dE, \beta=[E(E+2)]^{1/2}(E+1)^{-1}.
\end{eqnarray}
Then, after straightforward manipulation, equation (\ref{A3}) can be
expressed in the form,
\begin{equation}
  {\partial \over {\partial t}}\left(f(E,t)\right)=
  {\partial^2\over{\partial E^2}}\left[D(E)f(E,t)\right] -
  {\partial\over{\partial
E}}\left[\left(A(E)-\mid\dot{E_L}\mid\right)f(E,t)\right]-
  {f(E,t)\over T_{esc}} + S(E,t),
 \label{A7} 
\end{equation}
where
\begin{eqnarray}
D(E) &=& \beta^2D_p(p),\cr
A(E) &=& \frac{1}{p^2}\frac{d}{dp}\left(p^2\beta D_p(p)\right),\cr
\mid\dot{E_L}\mid &=& \beta \dot{\mathcal{E}}_L(p),\cr
S(E,t) &=& \frac{2\pi p^2}{\beta}Q(p,t).
\end{eqnarray}
The form of equation (\ref{A7}) is actually the Fokker-Planck form
of equation for a particle distribution function as originally
presented by {\it Chandrasekhar} [1943] for particles in stochastic motion.
Stochastic acceleration of electrons in solar flares has been treated
using different versions of (\ref{A7}), e.g., see {\it Ramaty} [1979],
{\it Petrosian} [1994], and {\it Park et al.} [1997].

\acknowledgments

This work is supported by the Natural Sciences and Engineering Research
Council of Canada under Grant A-0621. Additional support is acknowledged
from the Dean of Science, Memorial University of Newfoundland, as well as
NSF Grant ATM 97 29021 and NASA Grant NAG5 4680. Part of this paper was
written when D. Summers was Visiting Professor at the Radio Atmospheric
Science Center, Kyoto University, Japan. It is a pleasure to acknowledge
H. Matsumoto of Kyoto University for his generous hospitality and 
stimulating scientific discussions. We are also grateful to R. B. Horne 
and B. T. Tsurutani for helpful comments. 

{}

\end{article}
 
\clearpage

\begin{figure}
\caption{(Left) Schematic view in the magnetic equatorial plane of the
approximately circular (projected) drift path of relativistic electrons
in the inner magnetosphere. During storms these energetic electrons drift
(eastward) through regions of enhanced whistler-mode chorus and enhanced
electromagnetic ion cyclotron (EMIC) waves. 
(Right) Representation of the gyration about magnetic field lines and the
bounce motion of energetic electrons as they execute the approximately
circular drift path shown in (a).} 
\label{fig1}
\end{figure}

\begin{figure}
\caption{Diffusion coefficient $D_0$ given by (13) as a
function of the
turbulence spectral index $q$, for $2<q<4$. The upper, middle, and lower
diagrams correspond respectively to the locations $L=3, 4, 5$. The
background particle number density $N_0=10$ cm$^{-3}$. In each diagram,
curves are shown corresponding to the four indicated values of the wave
power $R$ (given by (9)) which correspond to the respective average wave
amplitudes $\Delta B=$75 pT, 100 pT, 300 pT, 1 nT.} 
\label{fig2}
\end{figure}

\begin{figure}
\caption{Diffusion coefficient ${\mathcal D}_0$ given by (14) as a
function of the average wave amplitude $\Delta B$ (pT), at each of the
locations $L=3, 4, 5$. The turbulence spectral index $q=5/3$, the
parameter $J_W=1$, and the background particle number density $N_0=10$
cm$^{-3}$.} 
\label{fig3}
\end{figure} 

\begin{figure}
\caption{Diffusion coefficient $D$ given by (3), (6), (13), (14) as a
function of the particle kinetic energy $E$, for the average wave
amplitude $\Delta B=1$ nT, and the indicated values of the spectral index
$q$. The upper, middle, and lower diagrams correspond respectively to the
locations $L=3$, 4, 5. The background particle number density $N_0=10$
cm$^{-3}$.}
\label{fig4}
\end{figure} 

\begin{figure}
\caption{
(Left) Steady state solutions $f(E)$ to the kinetic equation (1)
for the electron energy distribution function, for the indicated values of
$1/({\mathcal D}_0T_{esc})$, corresponding to different mean particle escape
times. The diffusion coefficient and systematic acceleration rate are
given by (6) and (7), with the diffusion parameter ${\mathcal D}_0$ defined
by (14); $q=5/3$, $J_W=1$, $L=3$, $\Delta B=75$ pT, $N_0=10$ cm$^{-3}$, and
${\mathcal D}_0=7.8\times 10^{-6}$ sec$^{-1}$. The particle source function
is given by (21), with $S_0=1.5\times 10^{-6}$ cm$^{-3}$ sec$^{-1}$ and
$\mu=20.44$.
(Right) Corresponding re-scaled plots of the solution curves on the left,
for comparison with relativistic Maxwellian energy distribution functions.
The dashed lines represent best fits with Maxwellian distributions in
accordance with the results given in Table 2.} 
\label{fig5} 
\end{figure} 

\begin{figure}
\caption{
(Left) Steady state solutions $f(E)$ to the kinetic equation (1)
for the electron energy distribution function, for the indicated values of
$1/(D_0T_{esc})$, corresponding to different mean particle escape
times. The diffusion coefficient and systematic acceleration rate are
given by (3) and (4), with the diffusion parameter $D_0$ defined
by (13); 
$q=2.5$, $L=6.6$, $\Delta B=800$ pT, $N_0=10$ cm$^{-3}$, and
$D_0=7.2\times 10^{-6}$ sec$^{-1}$. The particle source function
is given by (21), with $S_0=1.4\times 10^{-9}$ cm$^{-3}$ sec$^{-1}$ and
$\mu=20.44$.
(Right) Corresponding re-scaled plots of the solution curves on the left,
for comparison with relativistic Maxwellian energy distribution functions.
The dashed lines represent best fits with Maxwellian distributions in
accordance with the results given in Table 3.}
\label{fig6}
\end{figure} 

\begin{figure}
\caption{(Left) As in Figure 6 (Left), but for the parameters
$q=3$, $L=5$, $\Delta B=1$ nT, $N_0=10$ cm$^{-3}$, and
$D_0=6.5\times 10^{-6}$ sec$^{-1}$. The particle source function
is given by (21), with $S_0=1.9\times 10^{-8}$ cm$^{-3}$ sec$^{-1}$ and
$\mu=20.44$.
(Right) Corresponding re-scaled plots of the solution curves on the left,
for comparison with relativistic Maxwellian energy distribution functions.
The dashed lines represent best fits with Maxwellian distributions in
accordance with the results given in Table 4.
}
\label{fig7}
\end{figure} 

\clearpage

\begin{table}
\caption{Values of the magnetic field $B_0$ ($10^{-7}$ T), the parameters
$\alpha$ and $\beta_A$ given by (10), and the critical energy $E_c$ (keV)
given by (20), corresponding to the locations $L=3$, 4, $\cdots$, 9; the 
background
particle number density is $N_0=10$ cm$^{-3}$. Also given, corresponding
to the indicated values of the spectral index $q$, are typical values of the
average wave amplitude $\Delta B$ (pT) and the associated values of the
diffusion parameters ${\mathcal D}_0$ and $D_0$ in sec$^{-1} (\times 
10^{-6}$) required to produce a high energy ``hard" electron distribution 
after several days of sub-storm particle injection.\label{table1}
}

\begin{tabular}{ccccccccccc}

\hline

& & & & & 
\multicolumn{2}{c} {$q=5/3$} &
\multicolumn{2}{c} {$q=2.5$} &
\multicolumn{2}{c} {$q=3.0$} \\

\cline{6-7} \cline{8-9} \cline{10-11} 

$L$ & $B_0$ & $\alpha$ & $\beta_A$ &$E_c$ & 
$\Delta B$ & ${\mathcal D}_0$ &
$\Delta B$ & $D_0$ &
$\Delta B$ & $D_0$ \\ 

\hline

3 & 11.6 & 1.31   & 0.027  & 267 & 75   & 7.8 & 150  & 8.6 & 500  & 7.6 \\
4 & 4.85 & 0.23   & 0.011  & 57  & 200  & 9.8 & 300  & 9.3 & 700  & 6.2 \\
5 & 2.50 & 0.061  & 0.006  & 15  & 400  & 10  & 400  & 6.1 & 1000 & 6.5 \\
6 & 1.44 & 0.019  & 0.0032 & 4.8 & 600  & 7.3 & 600  & 5.7 & 1200 & 5.3 \\
7 & 0.91 & 0.0081 & 0.0021 & 2.1 & 900  & 7.0 & 800  & 5.4 & 1500 & 5.4 \\
8 & 0.61 & 0.0036 & 0.0014 & 0.9 & 1400 & 7.5 & 1200 & 6.6 & 1800 & 5.1 \\
9 & 0.43 & 0.0018 & 0.001  & 0.5 & 2000 & 7.6 & 1500 & 6.1 & 2200 & 5.4 \\

\hline 
\end{tabular}
\end{table}

\clearpage

\begin{table}
\caption{Results associated with the steady-state solutions shown in
Figure 5. Each line of the table corresponds to the particular value of
$T_{esc}$ (the mean particle escape time) indicated. Each solution is
fitted to a relativistic Maxwellian distribution by means of the linear fit
(26) in which the parameters $a$ and $b$ yield values for the number density 
$N_h$ and temperature $T_h$ of the distribution; $\chi^2$ measures the
goodness-of-fit. The time taken for the steady-state (equilibrium)
distribution to form is $T_{EQ}$.\label{table2}
}

\begin{tabular}{cccccccc}

\hline                                                          

$1/({\mathcal D}_0T_{esc})$ 	&	 
$T_{esc}$ (day)			& 
$a$ 				& 
$b (\times 10^{-3})$ 		& 
$\chi^2$ 			& 
$N_h(\times 10^{-4})$ cm$^{-3}$ & 
$T_h$ (keV) 			& 
$T_{EQ}$ (day) 			\\ 

\hline

0  & $\infty$ & -2.25 & -1.01 & 3.95  & 160  & 430  & 10 \\
1  & 1.5      & -3.41 & -1.11 & 0.30  & 8.9  & 390  & 8  \\
2  & 0.75     & -3.76 & -1.40 & 0.32  & 2.4  & 310  & 5  \\
5  & 0.3      & -3.88 & -2.17 & 0.11  & 0.8  & 190  & 3  \\
10 & 0.15     & -3.85 & -3.08 & 0.046 & 0.4  & 140  & 1  \\

\hline
\end{tabular}

\end{table}

\clearpage

\begin{table}
\caption{
As for Table 2, except the results are associated with the
steady-state solutions in Figure 6.\label{table3}
}
\begin{tabular}{cccccccc}

\hline                                                          

$1/(D_0T_{esc})$ 	&	 
$T_{esc}$ (day)			& 
$a$ 				& 
$b (\times 10^{-3})$ 		& 
$\chi^2$ 			& 
$N_h(\times 10^{-4})$ cm$^{-3}$ & 
$T_h$ (keV) 			& 
$T_{EQ}$ (day) 			\\ 

\hline

0  & $\infty$ & -3.47 & -0.58 & 0.77  & 3.8  & 750 & 11 \\
1  & 1.6      & -3.56 & -1.5  & 0.052 & 3.2  & 280 & 8  \\
2  & 0.8      & -3.74 & -2.0  & 0.034 & 1.2  & 220 & 5  \\
5  & 0.4      & -4.00 & -2.9  & 0.010 & 0.32 & 150 & 3  \\
10 & 0.2      & -4.17 & -3.9  & 0.034 & 0.13 & 110 & 1  \\

\hline

\end{tabular}

\end{table}

\clearpage

\begin{table}
\caption{
As for Table 2, except the results are associated with the
steady-state solutions in Figure 7.\label{table4}
}
\begin{tabular}{cccccccc}

\hline                                                          

$1/(D_0T_{esc})$ 	&	 
$T_{esc}$ (day)			& 
$a$ 				& 
$b (\times 10^{-3})$ 		& 
$\chi^2$ 			& 
$N_h(\times 10^{-4})$ cm$^{-3}$ & 
$T_h$ (keV) 			& 
$T_{EQ}$ (day) 			\\ 

\hline

0  & $\infty$ & -4.00 & -0.22 & 21 & 140  & 2000 & 12 \\
1  & 1.8    & -3.90 & -1.1  & 1.8  & 3.3  & 410  & 9  \\
2  & 0.9   & -4.02 & -1.5  & 0.88 & 1.2  & 290  & 5  \\
5  & 0.45    & -4.19 & -2.3  & 0.31 & 0.32 & 190  & 3  \\
10 & 0.2   & -4.30 & -3.3  & 0.12 & 0.12 & 130  & 1  \\

\hline
\end{tabular}

\end{table}

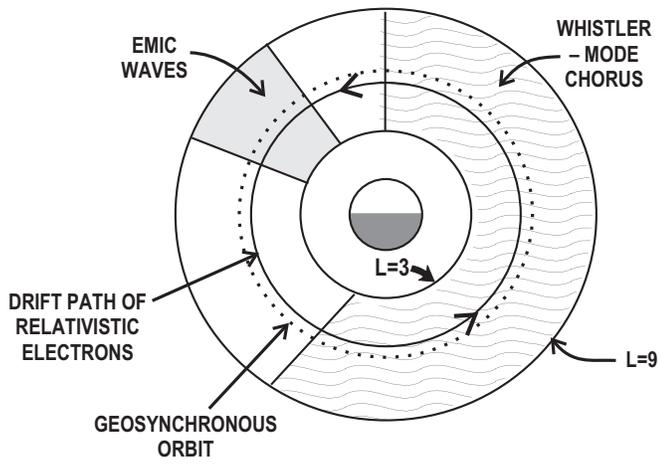
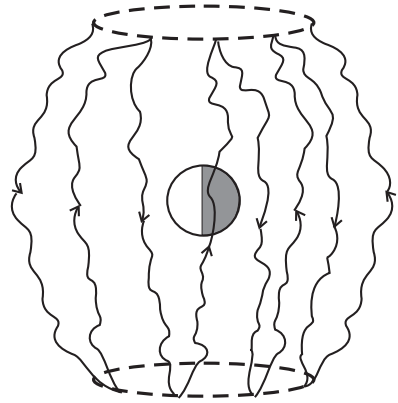
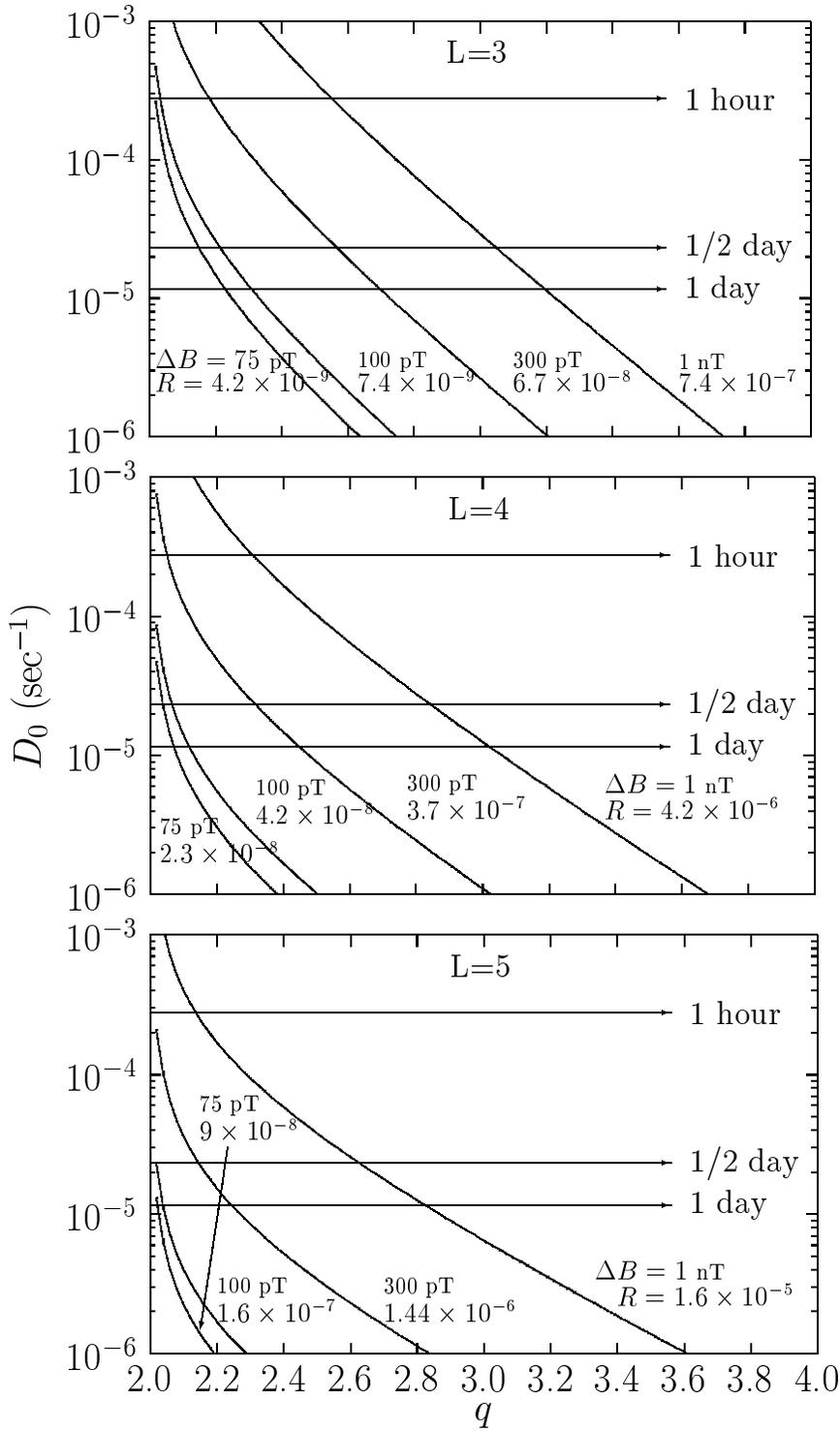
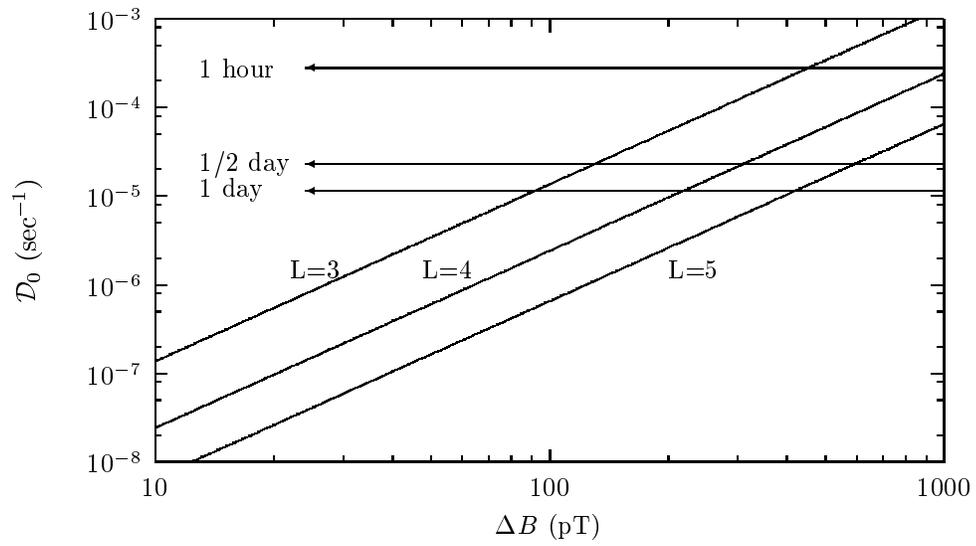
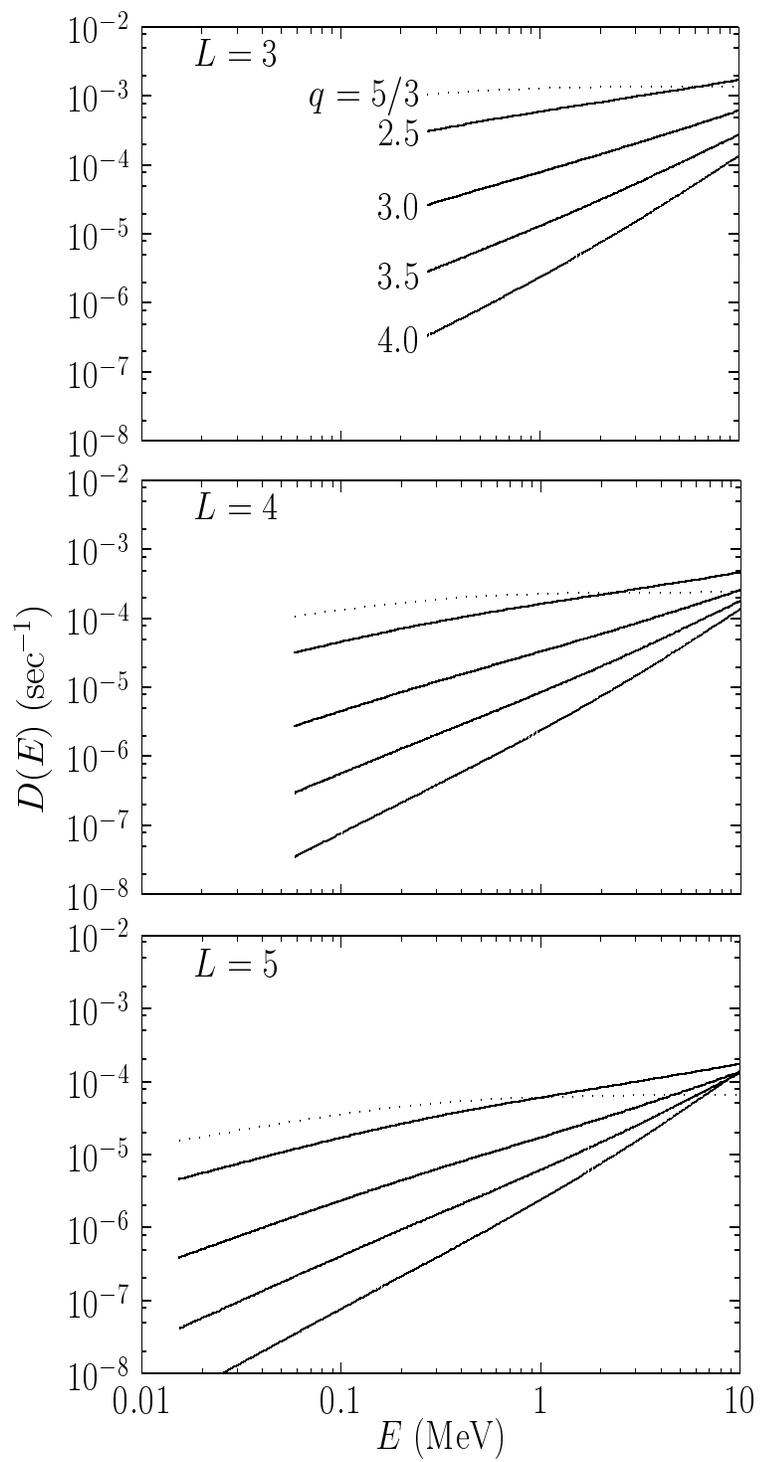
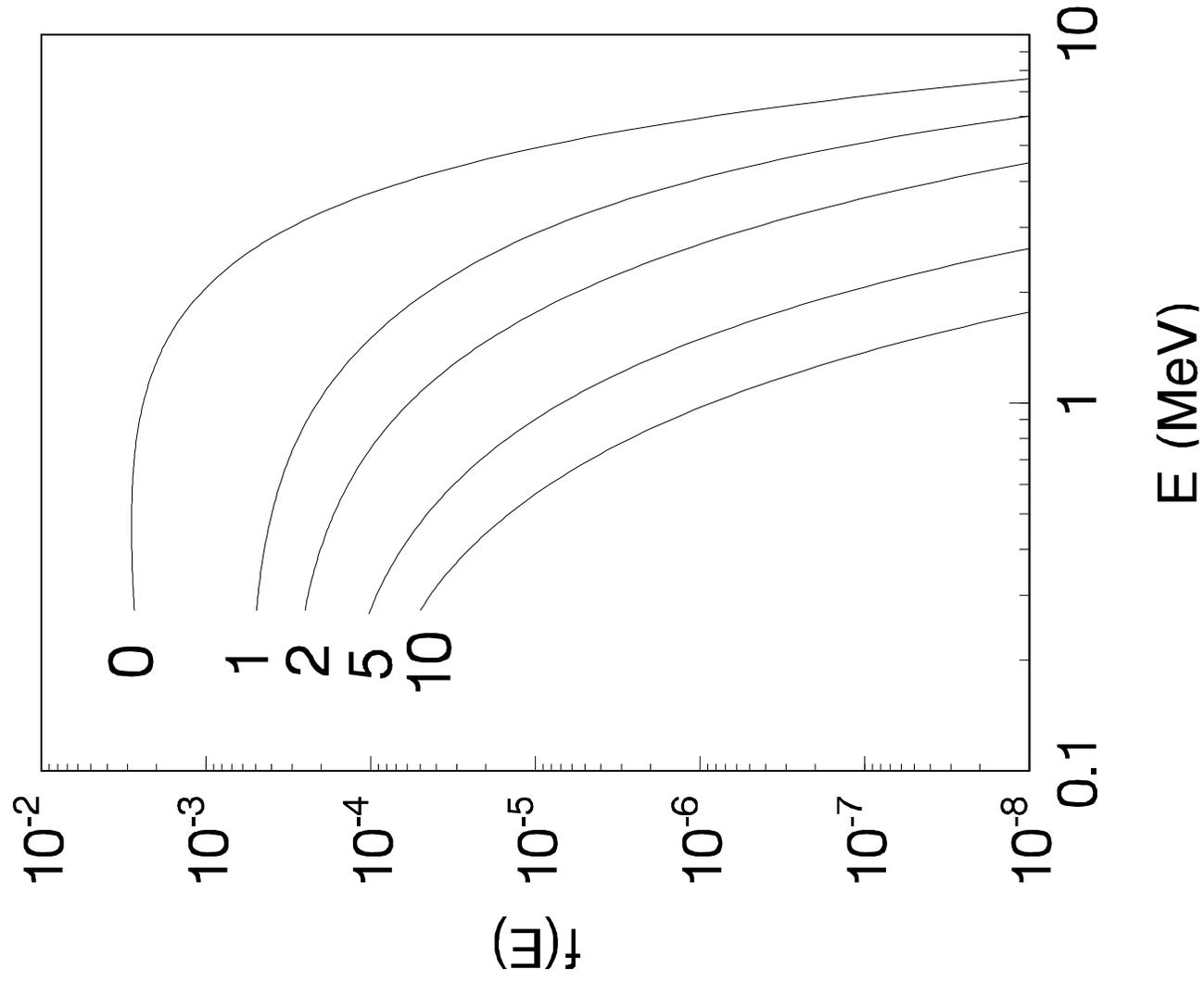
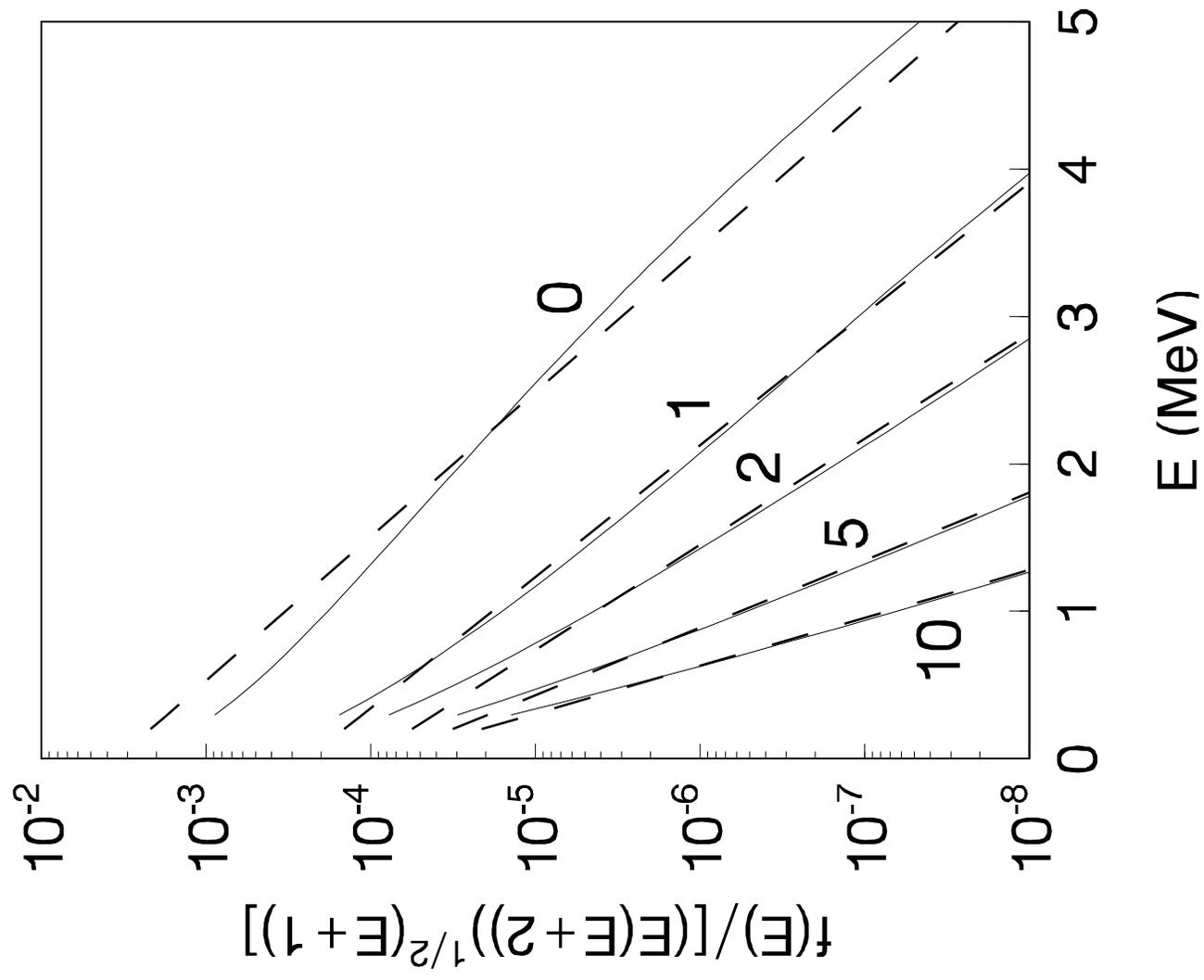
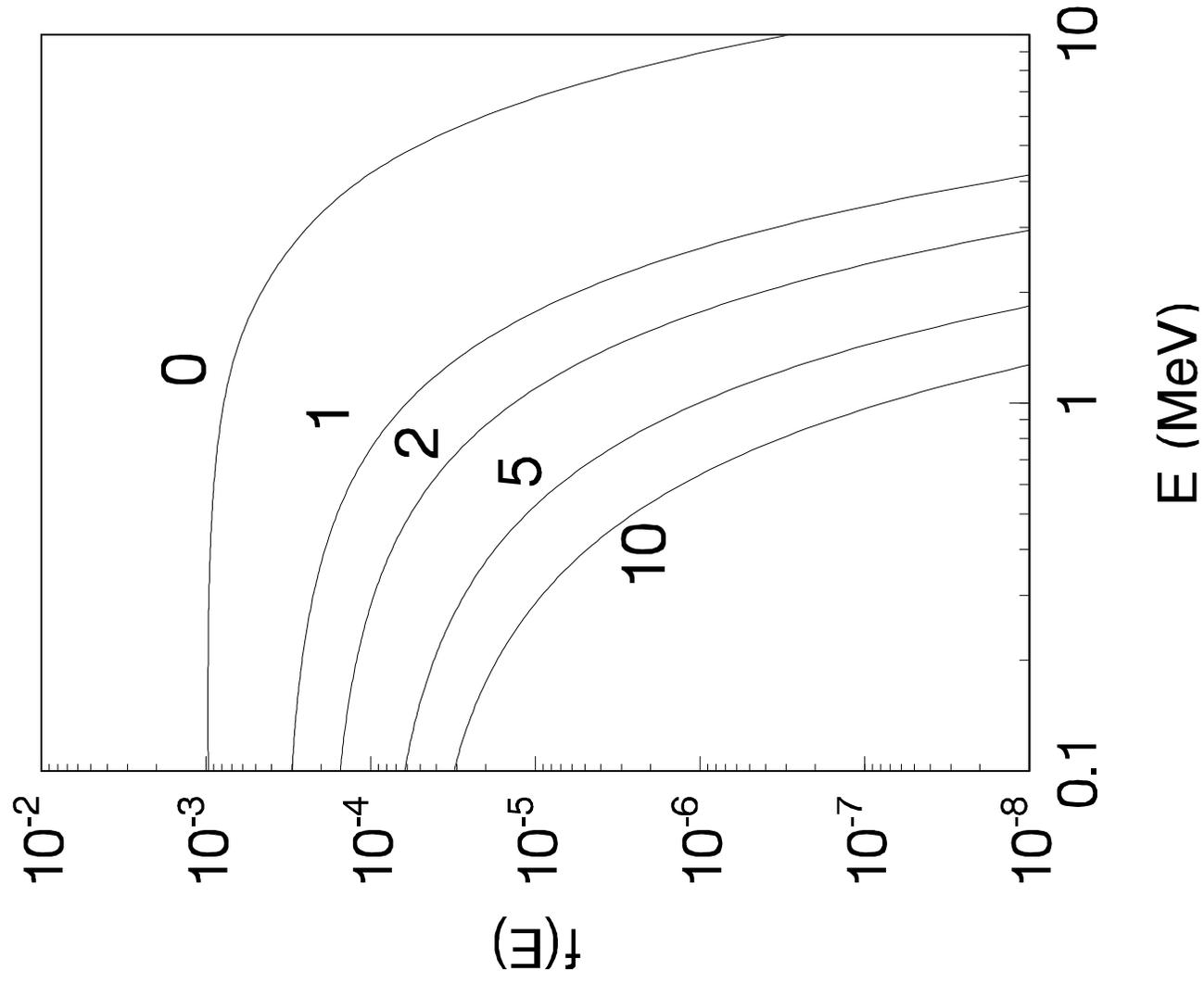
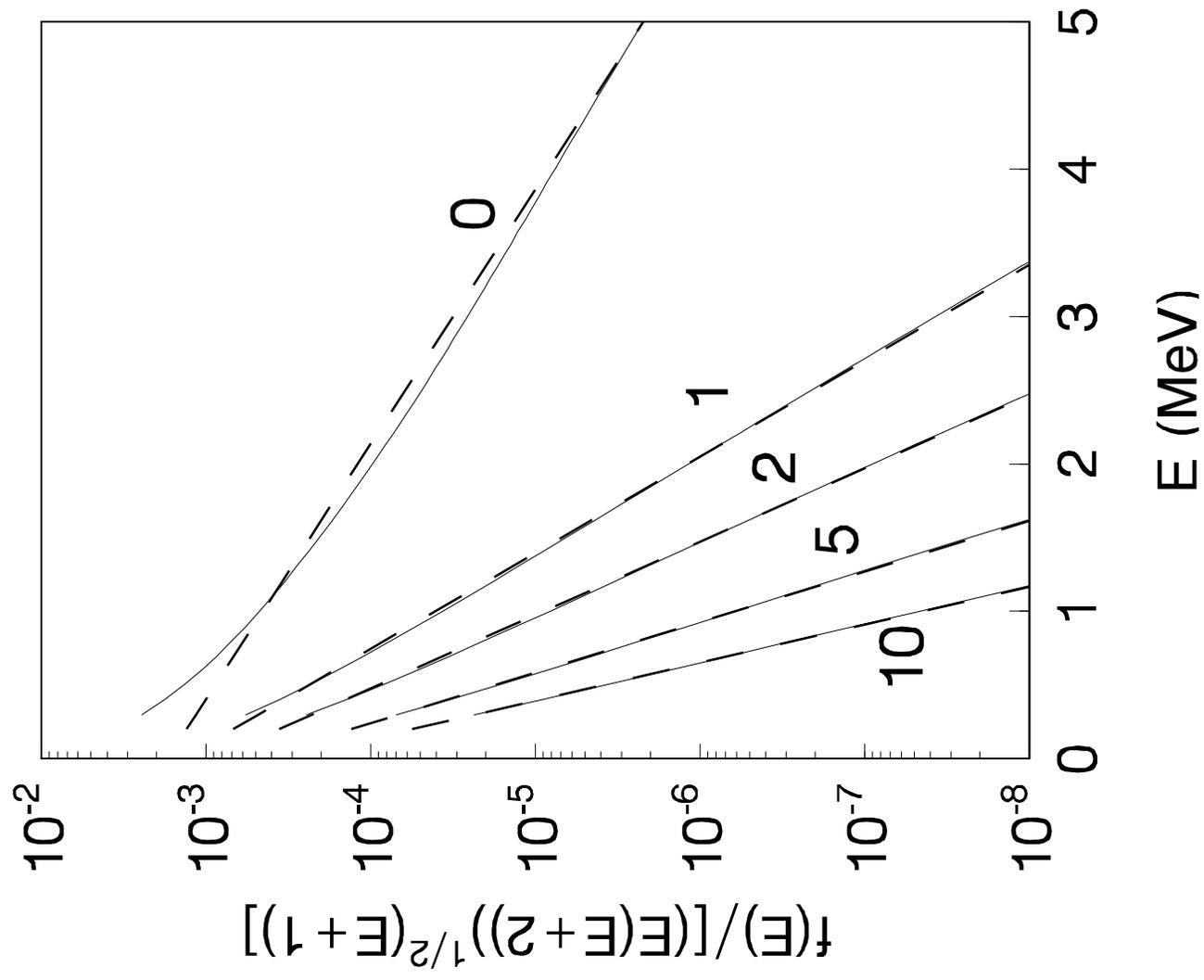
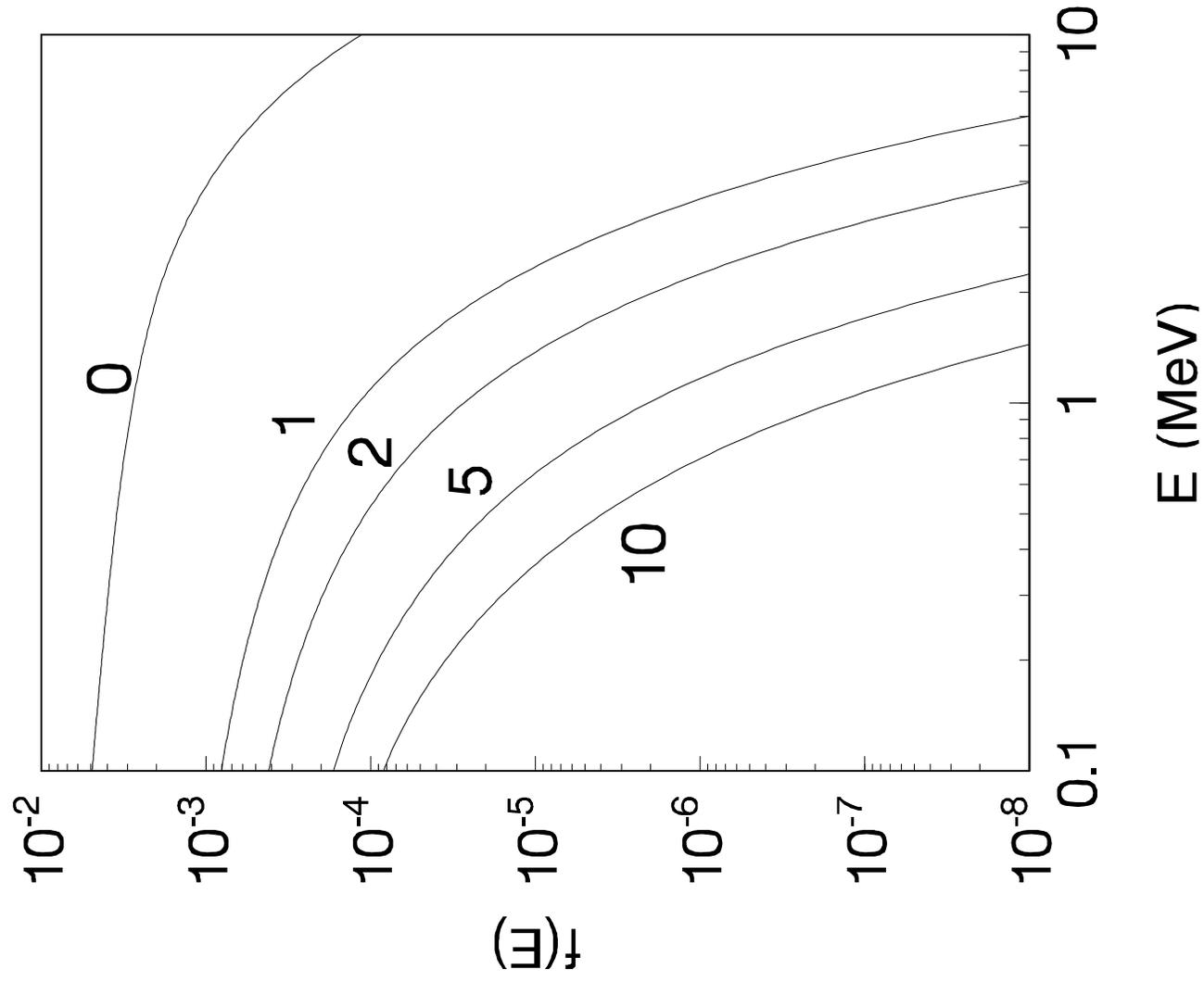
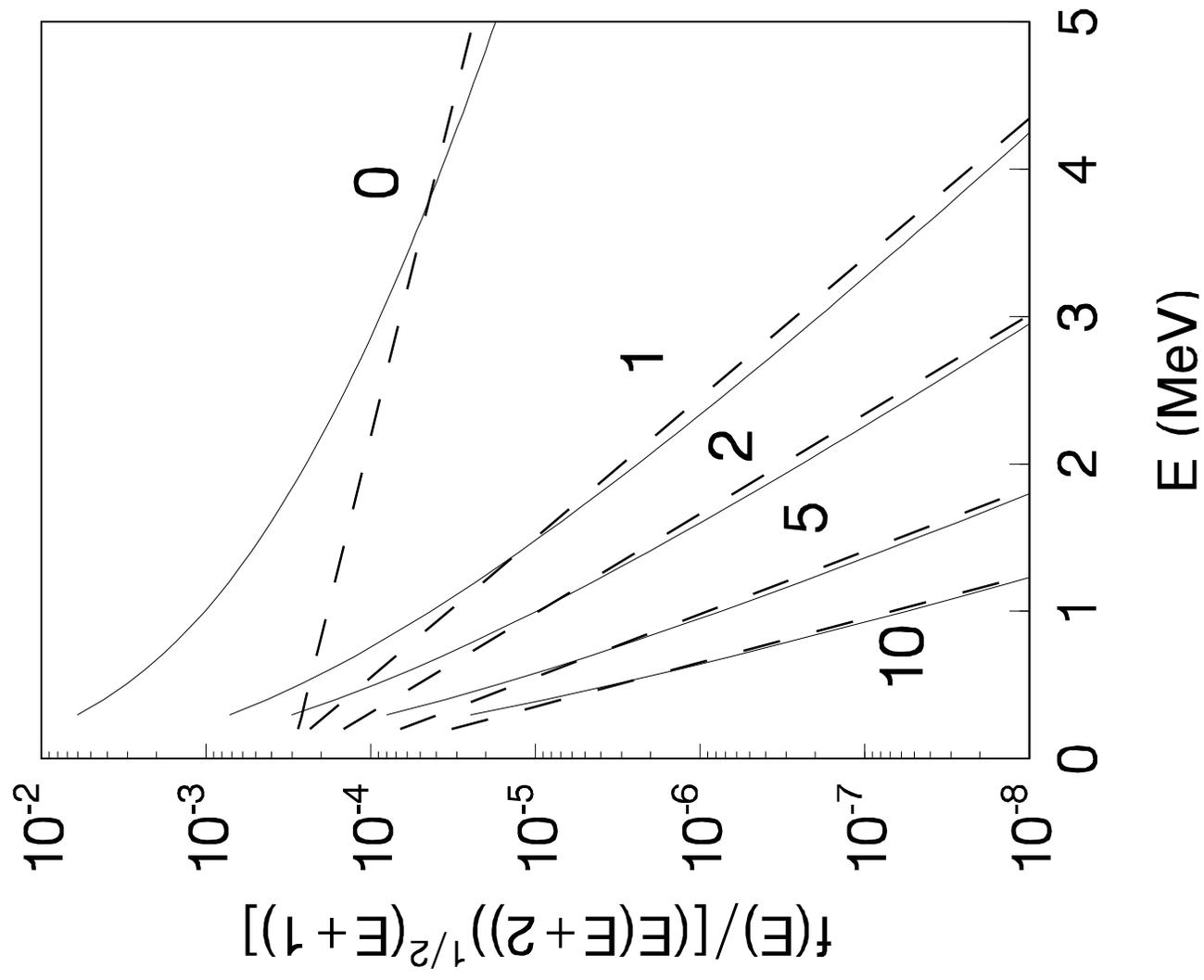
\end{document}